# STATE OF THE ART OF AGILE GOVERNANCE: A SYSTEMATIC REVIEW

Alexandre J. H. de O. Luna*[1,2], Philippe Kruchten[2], Marcello L. G. do E. Pedrosa[1], Humberto R. de Almeida Neto[1] and Hermano P. de Moura[1]

[1]Center of Informatics (CIn), Federal University of Pernambuco (UFPE), Recife, PE, Brazil.
[2]Department of Electrical and Computer Engineering (ECE), The University of British Columbia (UBC), Vancouver, BC, Canada.

*ABSTRACT*

***Context***: *Agility at the business level requires Information Technology (IT) environment flexible and customizable, as well as effective and responsive governance in order to deliver value faster, better, and cheaper to the business.* ***Objective***: *To understand better this context, our paper seeks to investigate how the domain of agile governance has evolved, as well as to derive implications for research and practice.* ***Method***: *We conducted a systematic review about the state of art of the agile governance up to and including 2013. Our search strategy identified 1992 studies in 10 databases, of which 167 had the potential to answer our research questions.* ***Results***: *We organized the studies into four major groups: software engineering, enterprise, manufacturing and multidisciplinary; classifying them into 16 emerging categories. As a result, the review provides a convergent definition for agile governance, six meta-principles, and a map of findings organized by topic and classified by relevance and convergence.* ***Conclusion***: *The found evidence lead us to believe that agile governance is a relatively new, wide and multidisciplinary area focused on organizational performance and competitiveness that needs to be more intensively studied. Finally, we made improvements and additions to the methodological approach for systematic reviews and qualitative studies.*

## 1. INTRODUCTION

Agility at the business level demands *capabilities*[1], such as flexibility, responsiveness and adaptability, which should be applied in combination with governance capabilities, such as strategic alignment ability, steering skills and dexterity to perform control; in order to achieve effective and responsive sense of coordination across multiple business units, especially in competitive environments. Under this context, the information and communication technologies (ICT or IT) are the link between the decision-making ability, the willingness strategic, and the competence to put into practice these tactics concretely. In fact, the design and maintenance of the IT systems for enterprise agility can be a challenge when the products and services must be compliant with several regulatory aspects (often needing to be audited). However, the establishment of the necessary management instruments and governance mechanism to fulfill this mission passes by the application of models and frameworks that many times have no guidance details of how to implement and deploy them (such as ITIL and COBIT, among others), affecting the organizational competitiveness [1], [2].

Before proceeding, it is important differentiate the well-known (1) specific agile approach widely held on organizations, such as agile software development or agile manufacturing; from the (2) agile governance approach proposed by this work. While the former has its influence limited to a localized result, usually few stages of the chain value [3] of the organization. Our proposal introduces the application of agility upon the system responsible for sense, respond and coordinate the entire organizational body: the governance (or steering) system. **Figure 1** depicts the difference between those approaches, in order to facilitate understanding: on part (A) we use an analogy that illustrate the anatomy of an organization as an human body; meanwhile the part (B) relates those approaches to the chain value concept proposed by Porter [3].

---

[1] "The power or ability to do something." [37]

We are also compelled to clarify the meaning of agility adopted by this work. In fact, we are adopting the agility definition proposed by Kruchten [S92][2] as: "*the ability of an organization to react to changes in its environment faster than the rate of these changes*". This definition uses the ultimate purpose or function of being agile for a business, unifying and standardizing *agile* and *lean* approaches as simply "*agile*", rather than defining agility by a labeled set of practices or by a set of properties defined in opposition to the Agile Manifesto approach [4]. Due of this simplified and objective approach, this will be the definition of agile adopted for this work.

To tell the truth, we recognize that while agility is focused on react rapidly to changes, lean is focused on combat the wastages. Although those approaches sometimes may seem confrontational if analyzed in its essence, we believe that the rational balance between those approaches can result in a unified "agile" approach that can achieve a better result than if they were applied separately, in consonance with Wang, Conboy and Cawley [S165].

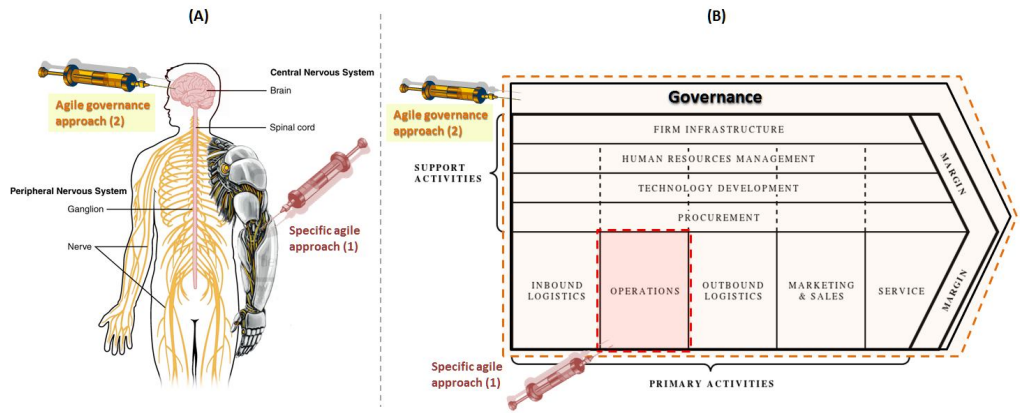

Figure 1. Organization's anatomy: an analogy. Source: Part (A), own elaboration; Part (B), adapted from [3].

Truth be told, when we look at the application of agility on governance it may seem like antagonist ideas (an *oxymoron*[3]) or counter intuitive, because governance denotes the idea of mechanisms, control, accountability and authority, while agility conveys the idea of informality, simplicity, experimentation, and for some observers (maybe) "almost anarchy". Nevertheless, if the goal of enterprise is to achieve business agility, it cannot be reached without commitment from all sectors of the organization, which in turn cannot be achieved without governance.

Based on those premises, arises as a relevant issue the understanding of the agile governance phenomena and the contexts in which they occur. Due once the agile governance phenomena are better understood in their essence, starting by its concept and application, as well as how it evolved over the time; become possible, in a second stage, map their constructs, mediators, moderators and disturbing factors from those phenomena in order to help organizations to achieve better results in their application: reducing cost and time, increasing the quality and success rate of their practice.

Hence, this paper reports on a systematic literature review carried out to map the state of art of agile governance (abbreviated by the acronym: SLR-AG), and it is part of a wider research conducted by the authors in order to identify combined application of **agile** *and* **governance** *capabilities* to improve business agility, as well as to extend the understanding of how these arrangements can help the organizations to attain greater enterprise agility and support its overall strategy.

---

[2] The citations highlighted as [S*] are studies included in this review, and their complete references are available at APPENDIX A.
[3] "A figure of speech in which apparently contradictory terms appear in conjunction." [37]

This article is structured as follows: In **Section 2**, is given an overview of agile governance, analyze the theoretical roots, and existent reviews. In **Section 3** we describe the methods applied and the methodological quality. **Section 4** presents the research results and characterizes the studies included on it; then the following subsections address research questions, some emerging contributions, and discuss strength of evidence; as well as consider indications for research and practice, and examine limitations of the review. **Section 5** concludes and affords recommendations for subsequent research on agile governance.

## 2. BACKGROUND

This section describes the field of agile governance, its root ideas, also how this domain have connection with other disciplines, and summarizes the critic thinking about agile governance. We present a summary of prior reviews of the agile literature, vindicate the necessity for this review, and expound the research questions that inspired the work.

### 2.1. Governance: the need to be agile

Governance is primarily related with **mechanisms** and **responsibilities** through which the **authority** is exercised, **decisions** are made and the **strategy** is **coordinated** and **steered** on the organizations, whether they are a country, an enterprise, a specific sector or a project. Calame and Talmant [5] introduce one of the best definitions to **governance**, that synthesizes the most important and distinctive aspects, while at the same time generalizing and universalizing the approach: *"Governance is the ability of human societies to equip themselves with systems of representation, institutions, processes and social structures, in order to they manage themselves, through a voluntary movement"*.

**Corporate governance** is the series of processes, policies, laws, customs and institutions affecting the way a corporation is conducted, administered or controlled, including the relationships between the distinct parties involved and the aims for which a society is governed [6]. **IT governance**, on the other hand, is defined by the IT Governance Institute as a subset of Corporate Governance, a discipline focused on information and communication technologies and their performance systems and risk management [7].

Dybå and Dingsøyr [8] posit that the Agile Methodologies have gained importance and add competitiveness and dynamism to the process of software development in the area of Software Engineering, through initiatives where the principles of communication and collaboration are crucial, as also stated in [S92] and [9]. Moreover, Dubinsky and Kruchten [S71], [S74] highlight that Software Development Governance (SDG) has emerged in the last few years to deal with establishing the structures, policies, controls, and measurements for communication and for decision rights, to ensure the success of software development organizations.

Recently, a proposal of **agile governance** has emerged. In 2007 Qumer [S54] presents the **first definition of agile governance** we found, focused on Agile Software Development. In an article published in 2009 about controlling and monitoring of product software companies, Cheng, Jansen and Remmers [S63] present the **second definition to agile governance** we found, focused on Software Development Governance (SDG). Additionally, in 2010 Luna, Costa, Moura and Novaes [S60] proposed a **third definition of agile governance**, focused on IT governance, resulting from the wide application of adapted principles and values of *Agile Software Development Manifesto* [4] to the conventional governance processes. In 2013, a **fourth definition for agile governance** was introduced by Luna, Kruchten and Moura [S150], as a result of perception of the multidisciplinary nature of the phenomena related to agile governance. All previously cited Agile Governance definitions are verbatim available in the **Table 7**. Hence, the concept of agile governance is gaining attention and evolving over the time as a meaning that is increasingly making sense in different contexts. In the sections that follow, we will dig into this issue gradually.

### 2.2. Summary of previous reviews

Based on the related work we looked for a previous systematic review related with the topic in many domains as follows. Dybå and Dingsøyr [8] point out some evidences about the application of agility beyond **Software Engineering** area, such as: agile manufacturing, lean development, new product development, interactive planning, maturing architectural design ideas and strategic management. These insights were very useful for our systematic review because they gave us some directions and helped us classify more accurately the findings of this research.

Wang, Lane, Conboy and Pikkarainen [S17] conducted a workshop identifying current agile gaps and areas for future research. From the sample of 161 papers published on XP conference until 2009, they classified ten of them as related to the emerging area of **business agility**, which was pointed out as one of **six emerging trends** that must be explored and studied and points the direction for where **agile research goes**. Although this approach cannot be considered a systematic review, it presented an **agile research topic map** that influenced the findings treatment of our systematic review.

In the **manufacturing industry**, Ramaa et al. [10] address the dearth of research on performance measurement systems and performance metrics of *supply chain network* by reviewing the contemporary literature, developing a systematic literature review. Their study lists more than 60 references for further study. They present four definitions for Performance Measurement of Supply Chain (PMSC), as well as a brief discussion about the evolution of this issue.

In the **IT governance** area, Qumer [S54] presents a summary of an exploratory review and analysis to identity the related concepts, key aspects and importance of IT governance, but he does not deepen the discussion. Correspondingly, Qumer proposes a conceptual "*agile responsibility, accountability and business value governance mode*l", for large agile software development environments. Likewise, Luna (2009) [11] conducted an exploratory review about the agile governance, using four electronic databases, found 75 references, trying to identify insights to propose a reference agile framework for implement and improve governance in organizations, called MAnGve, which is focused in the deployment process, as a "catalyst", accelerating the governance implementation.

Recently, Wang, Conboy and Cawley [S165] carried out an experience report analysis to provide a better understanding of **lean software development** approaches and how they are applied in agile software development. The findings of the study enrich our understanding of how lean can be applied in agile software development. The authors have identified *six types of lean application* in these experience reports and categorized them in a more systemic way: i) non-purposeful combination of agile and lean; ii) agile within, lean out-reach; iii) lean facilitating agile adoption; iv) lean within agile; v) from agile to lean; and, vi) synchronizing agile and lean.

However, we did not find systematic reviews in other areas of knowledge related with the combination of **agile** *capabilities* with **governance** *capabilities*. In other words, apparently, no systematic review about agile governance has been done yet. Therefore, there are no common understandings about the challenges that we must deal with, when examining the effectiveness of agile capabilities and governance capabilities, available for organizations and practitioners.

### 2.3. Objectives of this review

Preliminarily, no systematic review of agile governance has previously been found. The existing reviews that were presented in the preceding section are not systematic, or about this topic; neither covers the wide application of this field of study. In other words, this implies that executives, professionals, researchers and practitioners no have a unified reference to get an overview about this domain. The authors expect that this paper will be helpful to all of these groups, and that it will become clear which assertions on agile governance are sustained by scientific studies. This review aims to answer the subsequent research questions:

> *RQ1: What is the state of the art of agile governance in the world?*
>
> *RQ2: How the domain of agile governance has been evolved?*

In truth, to produce consistent findings with respect to agile governance, the review also ambitions to advance methodology for combining diverse study types, as well qualitative research, in systematic reviews of business agility interventions.

## 3. REVIEW METHOD

The work developed on this systematic review adopted as a methodological reference a combination from the following approaches: [8], [12]–[14]. This section describes the resulting approach.

### 3.1. Protocol development

A research protocol for the systematic review was developed by complying the guidelines, policies and procedures of the Kitchenham's Guidelines [13] and complemented by the Dybå's approach [8], as well as by the consultation with specialists on the topic and methods. Succinctly, our protocol establishes: i) the research questions; ii) search strategy; iii) inclusion, exclusion and quality criteria; iv) data extraction; and, v) methods of synthesis. The protocol is available in full version at the URL of the following reference [15].

## 3.2. Inclusion and exclusion criteria

In consonance with the research protocol, the studies were suitable for inclusion in the review if they offered evidences that helped to answer fully or partially at least one of the research questions, (see section 2.3). Due this work being a systematic review, whereby the authors were intended to identify the maximum of evidence to help them to set up a consistent view about the state of the art of agile governance, it was not defined a minimum quality threshold for exclusion of papers. In other words, the quality was a criterion of classification and assessment of strength of evidences, but it was not an elimination criterion (see section 3.6.1).

We included studies that addressed in their goals, hypothesis and applications, or analyzed, in their results, the combined application of agility and governance capabilities. Qualitative and quantitative studies published up to and including 2013 were included in the systematic review. We included only studies written in English.

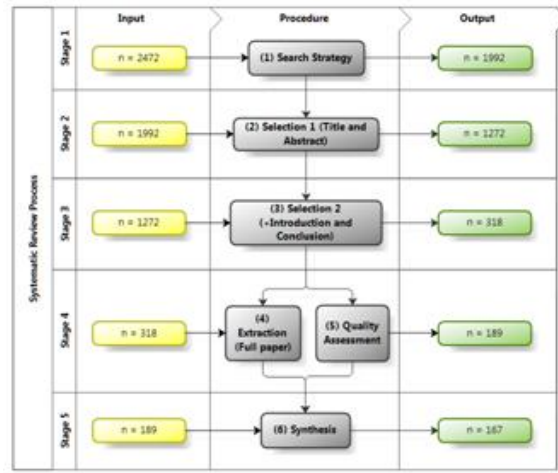

Figure 2. Systematic Review Process

We excluded studies if their focus were not: computer science, general theory of administration or general systems theory. In like manner, technical content that have not passed through a sieve of external review were excluded, such as: books, technical reports, dissertations, etc. In a similar vein, studies that are not complete articles were excluded, such as: extended abstract, keynotes, presentations, among others. Studies that expressing personal points of view or opinions were excluded, as well as articles that were in the areas of this research, but they clearly were not related to the research questions.

## 3.3. Data sources and search strategy

Our search strategy combined scientific electronic databases and the Google web search engine to identify and retrieve the entire population of publications that meets the eligibility criteria specified in section 3.2. We searched in the following electronic sources: (1) ACM Digital library; (2) Scirus; (3) IEEE Xplore Digital Library; (4) ISI Web of Science; (5) ScienceDirect; (6) Scopus; (7) SpringerLink; (8) Google Scholar; (9) Publish or Perish (POP); and, (10) Google PDF Documents*.

The **Figure 2** depicts the systematic review process and the quantity of papers identified and analyzed on each stage. At the **stage 1**, the titles, abstracts, and keywords in the aforementioned electronic databases (except Google PDF documents) were searched applying the following search terms: (1) *"agile governance"; (2) "lean governance"; (3) agile OR agility OR lean; (4) governance OR govern OR government OR "public administration"; (5) business OR enterprise OR corporate; (6)"service management" OR "information technology" OR "information and communication technologies" OR IT OR ICT.*

In addition, these terms were combined by applying the *boolean* operators, ''OR" and "AND", which implies that an article just had to include combinations of the terms to be retrieved, according the following "*search string*" (S):

   *S = (1) OR (2) OR ((3) AND ((4) OR (5) OR (6)))*

Eventually, we had to adjust the search string for the syntax of each electronic database, in other cases the searches had to be carried out in complementary steps (due to the limitation of the search interface of each database), though maintaining the same logic. Meanwhile, for the Google PDF Documents, due to the large amount of results, the *search string* was simplified to:

*S\* = (1) OR (2) filetype:PDF daterange:..2013-12-31*

Based on those ten electronic sources, the search strategy culminated in a total of 2472 ''results'' (or "hits") that included 1992 unduplicated citations.

### 3.4. Teamwork, tool and distributed research

At the same time, along this review was developed a software application (a tool) to support the research procedures. This software development was iterative and incremental, as well as the research database was hosted on a cloud, allowing the research was carried out in a geographically distributed way, by the internet. Therefore, for each stage, new interfaces and features were developed and made available for the researchers.

Table 1. Kappa coefficient and percentage of agreements.

| Stages | Kappa Pair 1 | Kappa Pair 2 | Kappa Pair 3 | Kappa Pair 4 | Kappa Stage | % Agreements |
|---|---|---|---|---|---|---|
| 2 | 0.88 | 0.57 | 0.28 | 0.23 | 0.61 | 80.6% |
| 3 | 0.74 | 0.70 | 0.39 | - | 0.65 | 86.6% |
| 4 | 0.78 | 0.88 | 0.72 | - | 0.80 | 91.1% |

During the stages 1 to 4 the team was arranged into groups composed by two researchers (or "pair"). At the stage 1 and 2, the researchers were organized into four pairs. Similarly, at the stages 3 and 4, the researchers were organized into three pairs. Meanwhile, at the stage 5, the work was developed individually, i.e., at this stage the set of research questions (RQs) were distributed among the researchers. Then they sought to answer the set of RQs based on the evidences found, and the result passed by a revision and cross-checking among them.

### 3.5. Citation management, retrieval, and inclusion decisions

At the **stage 1**, the electronic databases were randomly distributed among the pairs. Significant citations from this stage were entered into, complemented, organized and catalogued with the assist of *Mendeley*[4]. The results (or "hits") were then imported by the developed tool for a *research database* which was hosted on the cloud, where were registered the source of each citation, and the duplicity were removed by "title".

At the **stage 2**, the researchers of the same pair analyzed the title and abstract from the same "lot" of papers, that resulted from stage 1, to define their importance to this systematic review. At each stage, the researchers' retrieval decision, retrieval status, and eligibility decision were recorded on the research database. The detected agreement was (80.6%) and the Kappa coefficients of agreement [16] observed by each pair, for each stage are depicted in **Table 1**. The Kappa coefficient for stage 2 assessments was 0.61, which is typified as "substantial", in agreement with Landis and Koch [17].

At **stage 3**, the researchers analyzed (in addition to title, abstract and keywords) the *introduction* and *conclusion* for each paper from stage 2, to ascertain their pertinence to the systematic review. Whenever there was unsure whether a study complied to the screening criteria, based on the title, abstract, introduction and conclusion, the paper was included the next stage. The detected agreement was (86.6%), and the Kappa coefficient for stage 3 appraisal was 0.65, which is featured as "substantial" [17]. All disagreements were treated by discussion that comprised the two researchers from pair.

### 3.6. Full Analysis

The **stage 4** comprised the data extraction (DE) and quality assessment (QA) procedures that occurred simultaneously, but as a matter of clarity, they will be explained separately. The remaining studies were read entirely at least twice by each research of the same pair. After the second reading of the paper, each researcher (from the same pair) set his/her decision about exclusion or inclusion of each paper in the research tool. When the both researchers of the same pair finished the **individual analysis** the research tool pointed out if there were disagreements between them. All disagreements were treated by discussion that involved the two researchers from pair, the pair had to reach a consensus, otherwise, the quality assessment was carried out (see section: 3.6.1), the paper was extracted (see section: 3.6.2), and included for next stage. The observed agreements was (91.1%), and the Kappa coefficient for stage 4 was 0.80, which is typified as "substantial" [17]. Truly, under the pairs' point of view, at this stage we had two

---

[4]*Mendeley* is a desktop and web software for managing and sharing research papers, discovering research data and collaborating online [38].

"substantial" Kappa coefficients and other one "almost perfect", according the same criteria. This result was further evidence of the evolution of the team's learning in this process.

### 3.6.1. Quality Assessment

Each one of the remaining studies was assessed individually by each researcher of the same pair, in consonance with 13 criteria, in a procedure in which every criterion was graded on a scale composed by the following values: (2) when the criterion was plentiful or explicitly met; (1) when the criterion was partial or implicit met; (0) when the criterion was absent or not applicable. The quality assessment criteria adopted for these studies is briefly depicted in **Table 2.** The detailed criteria are disclosed in Appendix A. Quality assessment form of the research protocol [15]. These criteria were result from the analysis of methodological reference for appraising the quality of qualitative research [18], as well as by the principles and good practices established for driving empirical research in software engineering [8], [12]–[14].

**Table 2. Quality criteria.**

**Rigor Assessment Questions**
1. Is there a clear definition of the study objectives?
2. Is there a clear definition of the justifications of the study?
3. Is there a theoretical background about the topics of the study?
4. Is there a clear definition of the research question (RQ) and/or the hypothesis of the study?
5. Is there an adequate description of the context in which the research was carried out?
6. Are used and described appropriate data collection methods?
7. Is there an adequate description of the sample used and the methods for identifying and recruiting the sample?
8. Is there an adequate description of the methods used to analyze data and appropriate methods for ensuring the data analysis were grounded in the data?

**Credibility Assessment Questions**
9. Is provided by the study clearly answer or justification about RQ / hypothesis?
10. Is provided by the study clearly stated findings with credible results?

**Relevance Assessment Questions**
11. Is provided by the study a definition of Agile Governance (AG) or a definition some concept closely related with AG?
12. Is provided by the study justified conclusions?
13. Is provided by the study discussion about validity threats?

After the **quality individual assessment** step, when there were disagreements between the values attributed by the researchers on each pair about the same study, the tool pointed this out, and the pair had to reach a consensus, otherwise, a third researcher from the another pair (usually the more experienced) was invited to discuss the disagreements. Before the end of the stage all disagreements were solved by discussion.

### 3.6.2. Data extraction

Next, we decompose the research questions and identify the *constructs* for each one. The result of this step is available in **Table 3**. Along a third reading, data was extracted from every remaining study conforming to the predetermined extraction form (see Appendix B of the research protocol [15]).

Both the researchers from each pair extracted full data, from all studies, and then they discussed the data extracted during consensus meetings. Those evidences were gathered for posterior qualitative analysis of textual data. At the end of the **stage 4**, 1949 quotes were generated that helped answer fully or partially at least one of the research questions.

**Table 3. Constructs.**

**RQ1**
- Phenomena definition (meaning)
- Phenomena characterization
- State of theory and research
- Related fields: groups
- Context where take place: categories
- Overall trends

**RQ2**
- Genesis
- Evolution
- Perspectives

### 3.7. Synthesis of findings

Thereafter, during the **stage 5**, we use meta-ethnographic and qualitative meta-analysis methods, conforming to Noblit and Hare [19], to synthesize the data extracted from the studies. Initially the quotes were organized by research question and the coding of the constructs was complemented whenever it needed. We also carried out an open coding procedure to identify supplementary aspects of the relationship of the construct and the emerging categories. At same time, it was a manner of identifying relevant words, or classes of words, in the data and then tagging them accordingly.

The second step of the synthesis was to ascertain the main concepts and categories from each study, adopting the original author's terms. The process was carried out by organizing the key concepts in tabular form to permit comparison crosswise studies and the mutual interpretation of findings into higher-order meanings, following similar approach of **constant comparison** applied in qualitative data analysis [20]. When the researchers identified divergences in findings, they investigated whether these could be elucidated by the distinctions in methods or nature of the study setting [21].

## 4. RESULTS AND DISCUSSION

Unfortunately, our review did not identify any previous systematic review about agile governance. Due that, this work can be considered the first systematic review about the agile governance, in which we found 167 studies related directly or indirectly to this domain.

We will now discuss our results, beginning by addressing an **overview of the studies**, analyzing the **research methods** employed by them and discussing their **methodological quality**. After that, we will address the findings about **emerging groups and categories**, followed by **research questions** and closing with a consideration of the **state of art** of agile governance. Thereafter, the upcoming subsections discuss the **strength of evidence** of these findings, followed by the **implications of the findings for research and practice** and **emerging contributions**. Eventually, we discourse about the **limitations** of this systematic review.

### 4.1. Overview of studies

Concerning to the **nature of the research**: 101 studies (60.5%) were developed by researchers (in the *academy*) and 66 (39.5%) were carried out by practitioners or had the *industry* focus[5].

**Figure 3** presents the publication profile along the time, from the selected studies, grouped by year. Our review found no studies related with the issues of agile governance *prior to 1996*. In the same figure we can see three curves: the profile of publication for *academy* (A) and *industry* (I), as well as the *total of publication* (T) distributed along the years. Considering the phenomena as a nascent area we can approximate the data from T for a linear *trend line* (L), which equation is expressed in the **Figure 3**, presenting a coefficient of determination ($R^2 = 0.8029$). Based on this information, we can observe a *steady growth* of studies related with agile governance, reinforcing the idea of this area is in formation. In spite of the timeline publications can be expressed by a trend line, we cannot perform long-term forecasts because shall happen likely loss of linearity resulting from a significant event in the evolution of the phenomena in study.

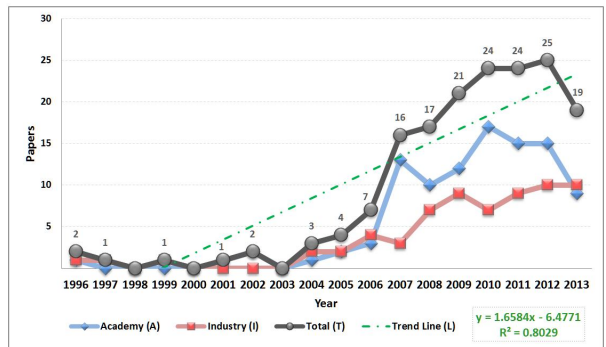

Figure 3. Review's Timeline: studies by year.

Each publication was further classified based on the **study type**: only 36 papers (21.6%) could be considered *empirical*[6] studies, which indicate a need for further studies conducted with more scientific rigor. The result from this analysis is listed in **Table 5**.

### 4.2. Research methods

The statistics of publication employing each research method are listed in **Table 4**. In the *mixed group* from **Table 4** were accumulated studies with more than one research method. At same time, Edmondson and McManus (2007) [22] advocate that the research design has to adapt itself to the current state of theory and research. Actually, they arrange this state into three types: *nascent*, *intermediate*, and *mature*. According to this classification, for agile governance, the large number of exploratory qualitative studies denotes that the studies on this area are still maturing, indicating that the present **state of theory and research** on methods is evidently *nascent*, which implies a demand for exploratory qualitative studies, originally open-ended data that have to be construed for meaning: what matches with the evidence profile found by this review.

Table 4. Studies by research method.

| Research Methods | Studies | % |
|---|---|---|
| Exploratory Analysis | 74 | 44.3% |
| Case Study | 36 | 21.6% |
| Mixed | 21 | 12.6% |
| Exploratory Literature Review | 15 | 9.0% |
| Survey | 11 | 6.6% |
| Experience Report | 4 | 2.4% |
| Workshop | 4 | 2.4% |
| Factor Analysis | 1 | 0.6% |
| Grounded Theory | 1 | 0.6% |
| **Total** | **167** | **100%** |

---

[5] That is, studies were performed in industry but sometimes were conducted in partnership with researchers or carried out by them.

[6] When the study demonstrated materiality and a coherent description about the methods applied (with consistent scientific rigor), conveying that it was based on evidence by means of experimentation or observation.

### 4.3. Methodological quality

Each study was assessed conforming to 13 quality criteria grounded on rigor, credibility and relevance, as described in section 3.6.1. Altogether, these 13 criteria give a measure of the range to which we can be trustful that a specific study's findings can generate a relevant contribution to this review. All included studies were graded as "full" or "partial" for the first screening criterion, which presenting the definition of the aims of the study. In general, this review often found that: methods were not quite explained; issues of bias, reliability, and validity were not always undertaken; and approaches of data collection and data analysis frequently were not well characterized.

Table 5. Studies by type.

| Study Type | Studies | % |
|---|---|---|
| Propositional | 70 | 41.9% |
| Expert opinion | 51 | 30.5% |
| Empirical | 36 | 21.6% |
| Application | 10 | 6.0% |
| **Total** | **167** | **100%** |

In complement, each study was classified according its quality score, in consequence of the sum of the 13 individual quality criteria, attributed by the researchers after consensus, before the end of stage 4. Accordingly, each of the 167 studies has a Quality Score between 0 and 26 points. For these 167 papers (*n*) were applied a statistical treatment of Pareto distribution [23] for the Quality Score (QS) calculated, and the results are depicted in the **Figure 4**.

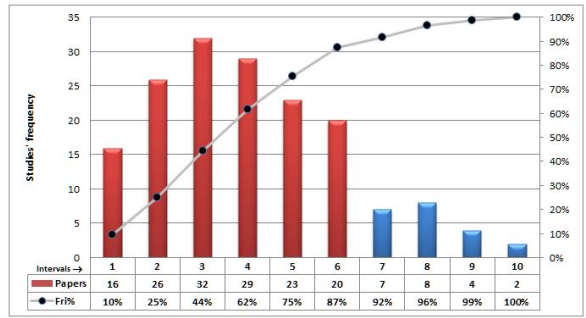

Figure 4. Histogram: Quality Score Pareto distribution.

The results denote that, only 72 studies (43.1%) have QS above the average ($QS_{average}$ = 10.7 points). Conjointly, 146 studies (87.4%) were placed into the first six frequency intervals with QS less than 16 points (intervals 1 to 6). In other words, just the 21 studies (12.6%) placed in the last four intervals of the **Figure 4** Figure present relative *quality significance* according the quality assessment criteria defined.

Further, each article was classified about the convergence of the study with the goals of this research, using the Likert Scale [24]. As a consequence, the result from this *convergence classification* was combined with the score of the *quality* assessment generating the **Figure 5,** which highlights the status of each paper about these two classifications: quality and convergence. In a similar vein, **Figure 5** can be analyzed in four quadrants, which were obtained using as a reference the middle point of each scale, arbitrarily defined. The *first quadrant* (Q1) joins the best of quality and relevance, positioning 34 studies (20.4%); while the *second quadrant* (Q2) present 6 studies (3.6%) with good quality score, but with less convergence to the aims of this review. On the other hand, the *third quadrant* (Q3) has 49 studies (29.3%) with low quality scores and little convergence; therewith the *fourth quadrant* (Q4) is represented for 78 studies (46.7%) with low quality scores despite to a good convergence to the goals of this review.

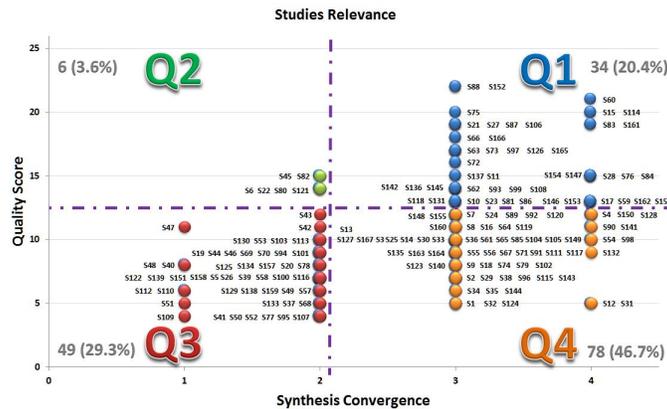

Figure 5. Scatter plot: Quality Score versus Synthesis Convergence.

Appropriately, 79.4% of the studies placed in Q1 were classified as empirical. Regarding the initiatives, this review found a rich set of contributions applicable in agile governance paradigm, from both academy and industry concerning to consistency and relevance about how the studies were conducted.

International Journal of Computer Science & Information Technology (IJCSIT) Vol 6, No 5, October 2014Although we prioritized the presentation of the findings of the studies positioned at the first quadrant (Q1) from the **Figure 5,** the remaining studies corroborate with the profile identified by studies positioned in Q1. Accordingly, during the discussion, we shall complement the results of Q1 with the findings of the studies from other quadrants to give more contexts for conclusion, when appropriate.

### 4.4. Emerging Groups and Categories

The set of evidences found was heterogeneous, leading the researchers to organize these studies into four major thematic groups: (1) software engineering, (2) enterprise, (3) manufacturing and (4) multidisciplinary. Following, we will introduce the studies contained in the four groups mentioned above. The 167 studies were classified according the approach of constant comparison applied in qualitative data analysis [20], following a bottom-up strategy: firstly trying identify emergent category, after trying to relate and group them, in a sequence of refinement cycles. Always as possible this classification was carried out according the authors point of view. In other words, when the authors were explicit about the category, their own classification was considered.

As a matter of fact, the authors were explicit about the category classification in 75 (44.9%) from the 167 papers selected. In 92 cases (55.1%), the category of the study was not explicit, and we had to compare the characteristics (content, objective and approach) of the paper with the categories previously identified to proceed the study classification. In other cases, the authors were not explicit about the category as well as the paper did not fit with any category previously identified: in these cases we had to propose a new category based on the paper characteristics. There were cases where the authors mentioned more than one category; in those cases we tried to identify which category was predominant, more consistent (or dense) in the paper content, to proceed with the study classification.

Table 6. Emerging (exclusive) Groups and Categories.

| Code | Description | Studies | % | Characteristic | Focus |
|---|---|---|---|---|---|
| G-1 | Software Engineering | 62 | 37.2% | Application of agile capabilities upon governance capabilities on Software Engineering | Software production |
| GASD | Governance for Agile Software Development | 23 | 13.8% | Application of governance capabilities on Agile Software Development | Agile Software Development |
| SOAG | SOA Governance | 20 | 12.0% | Application of agile capabilities upon governance capabilities on Service-oriented Architecture | SOA |
| SDG | Software Development Governance | 17 | 10.2% | Application of agile capabilities upon governance capabilities on Software Engineering | Software Development |
| SPIA | Software Process Improvement (SPI) Agility | 2 | 1.2% | Application of agile capabilities upon governance capabilities on Software Process Improvement | SPI |
| G-2 | Enterprise | 61 | 36.6% | Application of agile capabilities upon governance capabilities on Business Management or Public Administration | Enterprise as a whole |
| EA | Enterprise Architecture | 24 | 14.4% | Application of agile capabilities upon governance capabilities on Enterprise Architecture | EA |
| AE | Agile Enterprise | 16 | 9.6% | Application of agile capabilities upon governance capabilities on Enterprise's *modus operandi* (Business and Public Administration) | Organizational agility and responsiveness |
| e-Gov | e-Government | 10 | 6.0% | Application of agile capabilities upon governance capabilities related to solutions for electronic Government | e-Government |
| APA | Agile Public Administration | 9 | 5.4% | Application of agile capabilities upon governance capabilities on Public Administration | Government |
| l-Gov | Lean-Government | 2 | 1.2% | Application of lean capabilities upon governance capabilities on solutions for electronic Government | Government |
| G-3 | Manufacturing | 21 | 12.6% | Application of agile capabilities upon governance capabilities on Manufacturing | Manufacturing industry |
| AM | Agile Manufacturing | 12 | 7.2% | Application of agile capabilities upon governance capabilities on manufacturing process | Manufacturing |
| ASC | Agile Supply Chain | 8 | 4.8% | Application of agile capabilities upon governance capabilities on Supply Chain | Logistic |
| LM | Lean Manufacturing | 1 | 0.6% | Application of lean capabilities upon governance capabilities on Manufacturing | Manufacturing |
| G-4 | Multidisciplinary | 23 | 13.8% | Application of agile capabilities upon governance capabilities on many areas of knowledge | Holistic and wide approach |
| AITG | Agile IT Governance | 7 | 4.2% | Application of agile capabilities on IT Governance | IT Governance |
| APPG | Agile Projects and Portfolio Governance | 6 | 3.6% | Application of agile capabilities upon governance capabilities on Projects and Portfolio Management | Project |
| SG | Service Governance | 5 | 3.0% | Application of agile capabilities upon governance capabilities on Service Management | Service |
| LG | Lean Governance | 5 | 3.0% | Application of lean capabilities upon governance capabilities on IT Governance | IT Governance |
| | Total | 167 | 100% | - | - |

130

After the first cycle of classification for the studies, we started a classification refinement procedure, trying to review and confirm the classification from each cycle. In each cycle of this refinement procedure, each researcher revised the classification defined by the others. All discordances were treated by discussion that involved all the researchers who participated at this stage that had to reach a consensus. Any change in the classification, passed by the same procedure of revision and cross-checking among team researchers. The identification of emergent categories started with 34 original categories. After four refinement cycles, these categories were reduced to 16, sorted in four major groups, as depicted in **Table 6**.

Nonetheless, these groups are not isolated from each other. In truth, they have a strong relationship among each other,

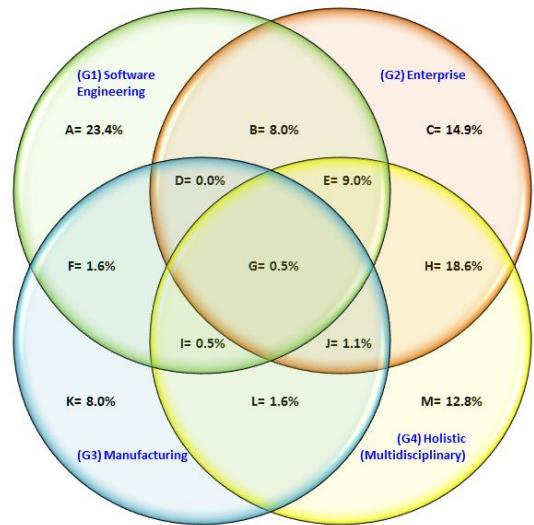

Figure 6. Emerging relationship (non-exclusive) between the major groups.

such as depicted in the **Figure 6**. Additionally, this figure is the result from a second *non-exclusive classification* about the relationship between the study characteristics and the four major groups, starting from the original and exclusive classification from the **Table 6**. In other words, adopting the same approach of revision and cross-checking applied in the exclusive classification, tags were applied for each study, trying to identify its relationship with the major groups, according the study characteristics.

Analyzing the data from **Table 6** and the description for each category, we could still try group some categories by focus (or core orientation). Under this approach the **G1** will not be affected. Nonetheless, on **G2**, the category **Lean-Government (l-Gov)** could be grouped in **e-Government (e-Gov)** category, due the second one might be a specific approach of the first one, and they have the same focus: application of lean or agile approaches in government, raising its representativeness for 12 studies (7.2%). Similarly, on **G3**, the category **Lean Manufacturing (LM)** may be grouped in the category **Agile Manufacturing (AM)**, because they have the same focus: manufacturing, changing its representativeness for 13 studies (7.8%). In like manner, on **G4**, the category **Lean Governance (LG)** might be grouped in **Agile IT Governance (AITG)**, due they have the same focus: IT governance, raising its representativeness for 12 studies (7.2%). Nevertheless, we preferred keep the original classification given by authors for final categorization of findings.

Regardless of the groups and categories identified by this review, in spite of the G1 has the most part of the isolated papers (23.4% from the 167 papers have no relationship with other group), the **Figure 6** denotes the relationship is denser between the G2 and G4 (sectors E ∪ G ∪ H ∪ J), due 29.3% from the selected papers are in these intersection regions. This finding implies the holistic nature from G4 and the wide approach of G2, inasmuch as the focus of the latter is the enterprise as a whole.

Comparing the values between the **Table 6** and the **Figure 6** is possible to identify a changing in the sort of representativeness of each group to this review. As a matter of fact, when we add the contribution of each sector from each group (even considering each intersection more than once, because some of them belongs to more than one group), the non-exclusive representativeness became: G2 (52%) > G4 (44%) > G1 (43%) > G3 (13%). This situation probably occurs due some studies originally classified exclusively according **Table 6** on the non-exclusive analysis carried out to generate the **Figure 6**, were classified also in other groups, in consonance with its characteristics. This phenomenon did not happen with the same intensity with the papers of G1 and G3, because their categories are more specific and have a better defined scope. Those evidences not only demonstrate the multidisciplinary nature of the agile governance phenomena, as well as they show a high degree of cohesion (intersection) among emerging groups.

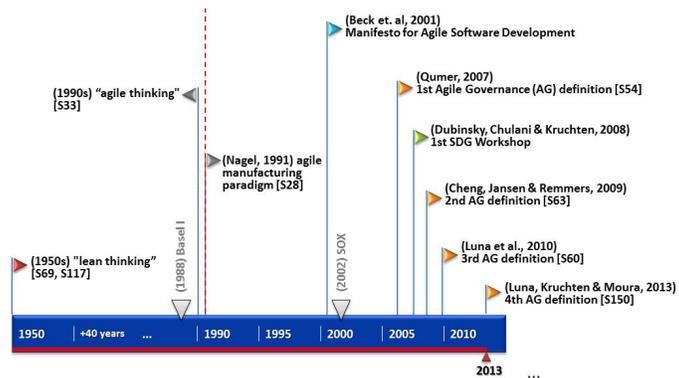

Figure 7. Agile Governance genesis timeline.

### 4.5. Genesis and evolution

We also develop the timeline depicted in the **Figure 7**, under the lens of the evidences found, in order to identify relevant aspects in the formation of agile governance field, as well as understand its *genesis* and *evolution*.

In retrospect, this review regained the recent history of agile governance that is intimately related with the "lean thinking" begun in the 1950s on Japanese industry [S117], [S69]. For a better understanding of the temporal relationship between these facts we plotted some marks that highlight the increasing of importance of global governance issues in the business, such as: the Basel I, the first of the three most important regulatory marks in bank market [25], as well as Sarbanes Oxley Act, the most important regulatory mark in stock market [26]. Our review found evidences that *agile philosophy* began at manufacturing industry [27] ten years before the Manifesto for Agile Software Development [4]. In reality, the "*agility thinking*" has entered in the literature in the early 1990s [S33]. However, as stated by Sun et al. [S28] just after the introduction of the agile manufacturing paradigm by Nagel [27], this concept began to attract significant attention from both the academy and industry.

Table 7. Agile Governance definitions

| Authors, Year | Focus | AG Definition |
|---|---|---|
| Qumer (2007) [S54] | Agile Software Development | *"an integrated agile governance involves lightweight, collaborative, communication-oriented, economical and evolving effective accountability framework, controls, processes, structures to maximize agile business value, by the strategic alignment of business-agile goals, performance and risk management"* |
| Cheng, Jansen and Remmers (2009) [S63] | Software Development Governance | *"the accountability and responsibility of management, adopting agile software development methods, and establishing measurement and control mechanisms in an agile environment".* |
| Luna, Costa, Moura and Novaes (2010) [S60] | IT Governance | *"is the process of defining and implementing the IT infrastructure that provides support to strategic business objectives of the organization, which is jointly owned by IT and the various business units and instructed to direct all involved in obtaining competitive differential strategic through the values and principles of the Agile Software Development Manifesto* [4] |
| Luna, Kruchten and Moura (2013) [S150] | Multidisciplinary | *"the 'means' by which strategic competitive advantages ought to be achieved and improved on the organizational environment, under an agile approach in order to deliver faster, better, and cheaper value to the business."* |

A good evidence to understand the genesis and evolution of the agile governance phenomena is analyze how the concept employed to describe it has been evolved over the time. Our review found only five[7] studies in which were encountered agile governance definitions, those studies and the verbatim definitions are depicted in **Table 7**. Chronologically, in 2007 agile governance was first conceptualized on **Agile Software Development** context. In this meantime, in 2008, was carried out the first **Workshop about Software Development Governance (SDG)**, led by Dubinsky, Chulani and Kruchten [28], as a landmark of the moment when this topic reached recognized significance in Software Engineering. Looking at the **Table** we can realize that the agile governance definition gradually had expanded its focus for **Software Development Governance (SDG)** in 2009, then to **IT Governance** in 2010, and reaching a **multidisciplinary** approach in 2013. This behavior is coherent as a domain that is taking shape, where the authors start to realize its amplitude and the relationships among the many contexts where the phenomena manifest themselves, broadly and holistically faceted and they try to cover their multidisciplinary scope.

---

[7] The study [S75] cites the same definition from [S54].

Furthermore, the **Figure 3** denotes a rapid growth process about total number of publications (T) found by this review after 2001, nearly doubling in the range 2001-2006, almost tripling in 2007. We believe that the behavior of the "T curve" agrees with the idea that this is a recent field of study in developing, as well as we rely on a trend of rapid growth of the publications related with this domain for the coming years. Regarding to the profile of publication for Academy (A) and Industry (I), in spite of the result expected for the industry profile to be usually try to experiment, test or/and apply the knowledge developed by academy, on this case we can see a little different behavior: we can observe that the Industry has followed the profile of publication of the Academy without significant lag, implying agile governance as a topic of practical and immediate applicability. The agile software development methods are phenomena that have a similar behavior in this aspect.

### 4.6. The state of art

The studies that handle over the adoption and introduction of agile methods on governance capabilities are still at an initial stage. Many of them were presented as a set of good intentions, but without a scientific rigor, which compromises their credibility and applicability. On the other hand, the big picture depicted by all of them do not give a unified view of ongoing practice, but offers a straightforward picture of experience and multiple fragmented findings. These issues is potentiate when we address aspects of agility in governance matters, a young and nascent area is **seven** years old, considering the publication of the first definition of agile governance by Qumer in 2007 [S54].

By means of the analysis of the publication timeline of the **Figure 3** we can realize two stages whereby the agile governance phenomena recently passed: (1) **the period until 2006:** in which we can see weak signals of agile governance as phenomena in formation expressed by few and intermittent publications; and, (2) **the period after 2006**: when the phenomena starting to take shape, with the first few published definitions, some categories emerging, the start of a language's construction, though still with many noises, distortions and ambiguities. As a consequence, we can imply that the **next great event on those phenomena** will be related with the *alignment of that language* to allow an adequate communication among the scholars and practitioners in an effective way. This episode will support the academy and industry to communicate and understand the phenomenon more clearly, and consequently admit achieving the necessary *fluency* in this area of knowledge in order to conduct it to a new baseline, accelerating its development.

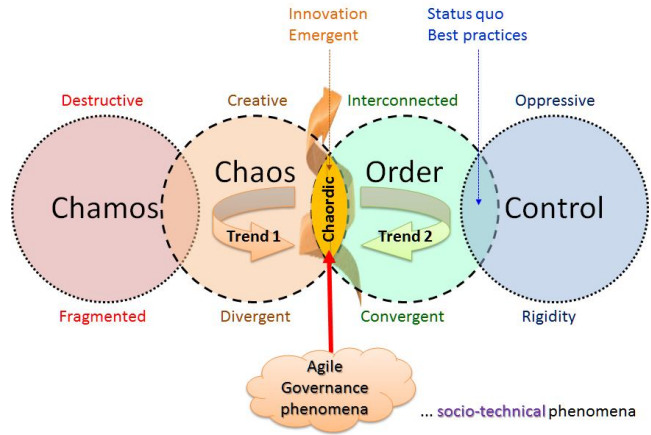

Figure 8. Positioning the agile governance phenomena.

Concerning to positioning of the phenomena, we can imply the agile governance as socio-technical phenomena positioned in a *chaordic* range between the innovation and emergent practices from agile (and lean) philosophy and the *status quo* of the best practices employed and demanded by the governance issues. The **Figure 8** depicts this phenomena's positioning proposal. The socio-technical nature of agile governance is substantiated due we are handling with the understanding of the intersections between technical and social aspects: considering people as agents of change in organizations, in contexts where technology is a key element. Actually, the *chaordic* philosophy was proposed by Dee Hock, the founder and CEO emeritus of VISA credit card association [29] as "*a system of organization that blends characteristics of chaos and order*" [30], as an harmonious and fertile business environment, whereas the duality of coexistence between chaos and order ends up becoming a propitious habitat for learning, transformation, growing, creativity and innovation.

In this context, agile governance inherits chaotic elements from the agile paradigm in which fit the agile and lean capabilities, whereas acquires ordering elements from the governance paradigm, including legal and regulatory aspects. This approach demystifies the discussion mentioned at the introduction that suggests agility and governance as alleged antagonistic ideas. At same time, it gives impulse for the consolidation of this concept as a creative and innovative balance between chaos and order, levering business achievements beyond the command-and-control conventional model.

In a complementary point of view, we can identify two *overall phenomena's trend movements*, in the agile governance paradigm, those are represented in **Figure 8**, based on the categories depicted in the **Table 6** : **(Trend 1)** mostly the categories related to G1 and G3 groups, *develop efforts to bring governance practices for their core issues* (respectively: Software Engineering and Manufacturing), leveraging existing agile culture in their environments; on the other hand, **(Trend 2)** mainly the categories comprehended in the G2 and G4 groups, *promote the endeavor of apply agile capabilities with governance capabilities for achieve better results in their core issues* (correspondingly: Enterprise and broad approach).

Although, these movements may seem contradictories, due they point out different (and apparently antagonist) directions into the same phenomena; those must be observed only as a point of beginning, due to the reality experienced in each context, to achieve the same results: apply agile and governance capabilities in combination. In other words, they are actually "*spin convergent*" (in a spiral movement), because the resulting vector of these two forces will reach the same result: unifying, adapting and accommodating particular components and specific issues in each area of application, to deliver value faster, better, and cheaper to the business.

The evidences found by this review lead us to realize the **urgent need for development of ontology for the agile governance paradigm**, as "*an explicit formal specifications of the terms in the domain and relations among them*" [31], organizing and relating the concepts, synonyms and adequate terms to express the ideas in a clear, straight and objective way.

If in one hand, there are a set of principles, practices and values from subjacent areas (as software engineering, manufacturing, government and business management) useful to apply in agile governance context. On the other hand, **these set of constructs are not organized and systematized for direct and immediate application**: they need be *translated* and *adapted* for each context. Truly, the available knowledge has to be *suitable* for the broad context of this domain, and our review did not found a guide, model or framework that can help to apply this knowledge in a systematic and adaptive manner. We believe that all those set of knowledge should be organized, connected and systematized in some kind of *conceptual framework* or *theory* [32].

### 4.7. Emerging contribution

During the synthesis and refinement process of the review's findings, our perception and sensibility were shaped by the wealth of detail found about agile governance in the different areas of specialization identified guiding us to generate further contribution in order to provide some initial impulse to help the development of this field.

The emerging evidence of this review lead us to believe that ***agile governance* can be *broad* and *holistically* defined**, as:

> "is the ability[8] of human societies to sense, adapt and respond rapidly and sustainably to changes in its environment, by means of the coordinated combination of agile and lean capabilities with governance capabilities, in order to deliver value[9] faster, better, and cheaper to their core business."

When we mentioned the term "**human societies**", we try to encompass any kind of organizations, such as: companies in any industry, non-profit institutions, as well as governments in any level or conjunction (cities, provinces, countries, or even governments associations, e.g. The United Nations).

In turn, "**core business**" is the *raison d'être* of any organization, the cause of its existence. When the organization identifies its customers and recognizes which kind of benefit or value (by means of products and services) they are delivering to customers in order to achieve its institutional mission, they are addressing their core business. As a matter of fact, this concept can be applied for any kind of organization, for instance: in case of a **company** may be the target activity to

---

[8] "A natural or *acquired* skill or talent."[39].

[9] "An informal term that includes all forms of value that determine the health and well-being of the firm in the long run." [41]

International Journal of Computer Science & Information Technology (IJCSIT) Vol 6, No 5, October 2014

achieve profit, for a **NGO**[10] might be a variety of service and humanitarian functions, concerning to **governments** should be initiatives to accomplish the welfare of its citizens.

Gradually, *business agility* has become an expression that is not restricted to the universe of for-profit organizations. In consonance with the proposed definition, we distill a *new definition* to **business agility** as:

"*the ability to deliver value faster, better, and cheaper to the core business*".

This new agile governance definition is being presented in order to be comprehensive enough to cover all areas identified by this research, at the same time that it is still specific enough to be useful and applicable in each of these contexts, avoiding being another definition disconnected from the multidisciplinary nature of this wide field of study.

In spite of many of scholars can criticize the absent of the "process" concept on the aforementioned definition, we would anticipate in saying that agile governance is related much more to behavior and practice than anything else. Even because processes and procedures are already well established in governance context, and they "need to be followed", many of them needing to be audited [33], or regulated by laws [26], or else certified as international standards [S90].

At this point we would like to clarify that agile governance do not come replace the conventional models, frameworks and methods, such as ITIL [2], COBIT [1], among others. Our proposal is just come shed a fresh look about governance, bringing enablers elements from agile philosophy to extend it for a more resilient and flexible paradigm. Actually, all knowledge relevant and useful existing related to governance topic have to be organized in some kind of dynamic referential[11] repository which we will denominate conceptually of *Governance Body of Knowledge (GBOK)*, which it should be organized systematically, fluidly and flexibly, as well as it does not end in the models cited in this work or known at the time of this publication. In fact, it must be complemented, organized, and must have a scope and boundaries better defined in future works.

At same time, the synthesis of our findings when combined with the approach of agile methods on governance capabilities lead us to propose the following six **meta-principles** for agile governance, in order to guide future researches and, especially, to drive the practices:

1. **Good enough governance:** "*The level of governance must always be adapted according to the organizational context*". The level of governance required to achieve business agility must be balanced, and adjusted when needed, taking into account the particular conditions, and timing[12] of each organization. This meta-principle should lead the practitioners and researchers to reflect and consider the constraints experienced by each organization, without jeopardize the regulatory aspects or market rules. In other words, it can be accomplished respecting the particularities of each environment. For instance, something that is good for an organization can be too much for other, at least on a specific time frame. The question remains: is it worth paying for this "extra"? Taking for example the COBIT 5 framework [33], which has 37 processes, and 17 enterprise goals. Shall these processes and goals be applied in any cases? In any kind of organizations?

2. **Business-driven:** "*The business must be the reason for every decision and action*". Decisions of any nature, in any organization instance, must be driven by and for the business. In other words, all decisions in any business unit, from entire organization (including its conjunctions and specific sectors) must be made taking into account the business strategy. People have to think each decision, design and approach to satisfy business requirements and priorities. Teams should create a broad culture that can influence the collective behavior in whole enterprise, in order to give rise to a cohesive organizational awareness. As a result of the alignment between the business layer and the governance layer, the connections among each unit of the entire organization, may work as a symbiotic relationship. This leads the organization to increase flexibility and to reduce the turnaround times when the business demands quick adapting of the infrastructure to its needs.

---

[10] "A non-governmental organization (NGO) is any non-profit, voluntary citizens' group which is organized on a local, national or international level." [40]

[11] Due many components of GBOK are proprietary models, guides or frameworks, the initial idea can be create an *index of references* relating those components to relevant aspects of agile governance, or even to the groups and categories identified by this review.

[12] "The selecting of the best time for doing or saying something in order to achieve the desired effect." [39]

3. **Human focused:** "*People must feel valued and incentivized to participate creatively*". People have to be valued as a key element of change and the driving force in organizations, as well as they must be encouraged to contribute creatively to the business aims. In organizations there are people who perform, control and decide about the processes, in so far there must be leaders that aim to create value in the company by means of getting the best from people, motivating them strategically, to obtain the need engagement to the business. Nonetheless, mostly the prevailing methods and tools of governance still are concentrated on structures and processes. The necessity to understand people as an essential and creative component of the structures and processes is a *critical success factor* for governance initiatives. At the same time, the creation of effective mechanisms to incentive and support the relationship, communication and collaboration among people is imperative.

4. **Based on quick wins:** "*The quick wins have to be celebrated and used to get more impulse and results*". The quick wins achieved by team must be celebrated with the same intensity and seriousness with which the problems are addressed and solved, as well as its impulse must be used consciously to get more results. The quick wins seek an accumulation of small impulses which, together in the same direction, are reflected in the medium and long term great acceleration to the enterprise. This evolution must be continuously monitored and adjusted. The maturity achieved by the team reflects on "less jerky movements", less breakage and waste, as well as greater coordination between the parties involved (people, business units, etc.). The "positive energy" coming from these victories must be used consciously in the feedback and motivation to the team to continue development of the governance initiatives and, therefore, should be valued.

5. **Systematic and adaptive approach:** "*The teams must develop the intrinsic ability to systematically handle change*". They should adopt a systematic and adaptive approach (adjusting the direction in line with the moment experienced by the organization). The teams and business units should seek to work as organisms adaptive rather than predictive ones. In other words, they should consider the change as natural component of the business environment, trying to adapt themselves to new factors arising from the development of their environments, as well as the business needs, rather than try to analyze previously all that can happen during each time box.

6. **Simple design and continuous refinement:** "*Teams must deliver fast, and must be always improving.*" That is to say, they must choose always the simpler and feasible alternative to the solutions design, one that can be improved with the least possible waste at the earliest opportunity. The idea is to adopt simple design and to improve it as soon as possible, instead of a slow start, trying to establish a balance between the agile and lean approach. The architecture of the solutions should always be focused on streamlining between the desired results and the resources currently available. In other words, it is better to do something simple that generate results immediately, and pay a little more to improve it at the first opportunity (by means of a possible rework), than doing something complicated with a high cost of time and other resources, and end up losing the timing of the change in the business.

In fact, in these meta-principles, "*team*" is a generic word that can be applied for several complementary connotations in organizational context, such as: technical people, business people, and even the steering committee. Besides, the adoption of the Greek prefix "meta" to characterize them is due to our having designed these principles to provide a way of thinking across the disciplines that compose the agile governance phenomena, trying to cover their broad nature. Also we should clarify that the these meta-principles were shaped under the lens of the principles analysis method proposed by Séguin et al. [34], properly adapted to the phenomena in study in this review.

### 4.8. Strength of evidence

Using an approach similar to Dybå and Dingsøyr (2008), we adopted the GRADE working group definitions to assess the entire strength of our review's evidences [35]. In relation to **study design,** were identified only eight studies where the authors argue that they have developed some kind of experiment, while the remnants primary studies were observational. Regarding to the **quality** of the studies, truly methods were not well characterized, in general; issues of validity, bias, and confidence were not always undertaken; and approaches of data collection and data analysis frequently were not well reported. Concerning to **consistency**[13] of the studies, were not

---

[13] "The similarity of estimates of effect across studies." [8]

International Journal of Computer Science & Information Technology (IJCSIT) Vol 6, No 5, October 2014found significant differences of alignment among them. In spite of studies present some concepts described in different way, or use some different words to describe same concepts, this situation represent, in our opinion, the **idea of incompleteness, but not the idea of inconsistency**. In relation to **directness**[14], there are enough consistence and coherence between the studies selected by this review, but they present a fragmented view about the domain of agile governance.

Analyzing these four components of study, we identified that the *strength of the evidence* in the present review concerning to the state of art of agile governance is **low**. Thence, any estimation of effect that is grounded on evidence of agile governance from current research **has low certainty**. This evaluation is consistent with the nature of a nascent field and with the fact of this review be pioneer in an unexplored area, quite lacking of studies.

### 4.9. Implications for research and practice

Several inferences for research and practice can be derived from our systematic review. However, companies and practitioners should use the findings of this review with critically discernment, in order to identify resemblance and discrepancy between the studies presented and their own reality. For **research**, this review demonstrates a clear necessity for studies with more scientific rigor, according the quality assessment developed in section 4.3. Our review found only 36 studies (21.6%) that were empirically conducted, which indicates a need for further empirical studies.

Our review confirms that agile governance has a wide spectrum of interest for executives from any business area, professionals, researchers and practitioners by treating, in essence, aspects such as: organizational performance and competitiveness, as well as it can be verified by the categories and major groups that emerged from these research findings. We believe that *researchers* and *practitioners* in agile governance should cooperate to define a common research agenda, for the sake of enlarge the usefulness and suitability of the research **for industry** and to produce an enough quantity of studies of great quality on subtopics associated to this field. We recognize that is outside the sphere of this paper propose such agenda, but we expect that the synthesis of research conferred herein may give the inspiration to conceive one.

Likewise, we invite enterprises to engage in research projects in the future, with the view to address research aims that are significant for the industry. Truly, action research is a contributive and convergent way of systematize cooperation between researchers and industry that would be immensely positive for a flourishing field such as agile governance. In retrospect, these findings represent a great opportunity for new research and practice on this domain. At same time, they might represent an obstacle to be overcome for a most significant advance in the production of this field, for industry and academy.

### 4.10.   Limitations of this review

Usually, the major limitations of a review are biases in the *publications selection* and *imprecision in data extraction procedures* [13]. In furtherance, to aid to assure that our selection process was unbiased, we elaborated a research protocol beforehand that delimited the research questions. Applying these questions as reference, keywords were identified and search terms, which would allow us to recognize the appropriate literature, were developed and tested.

We do recognize that the keywords, generally, are not patterned, much less in a multidisciplinary context such as agile governance. Besides, the subtlety of its application can be consequence from both knowledge domain and language specificity. Accordingly, by cause of our decision about the search strings and keywords, as well as the syntax of the **search mechanism from each electronic database**, there is a chance that pertinent studies were omitted.

Toward to minimize **selection bias**, every stage of the review process had a pilot practice. Special attention was given to the **search strategy** along with citation management procedure (stage 1 from **Figure 2** ) in favor of clarify vulnerabilities and refine the selection procedure. Strictly, to further insure the impartial selection of studies, a multistage process was applied. Furthermore, this process (depicted in **Figure 2**) involved two researchers (or "pair") at the stages 1 to 4 who recorded the justifications for inclusion/exclusion at each step, as detailed in Section 3 and as suggested by Kitchenham et al. [13]. Indeed, none study was excluded without a consensus in the pair's analysis. Even in relation to the low Kappa coefficient in the initial stages of this review (see **Table 1**), the authors believe that it does not impact on the credibility of the decisions (inclusion or exclusion) because, whenever the researchers do not reached a consensus, the paper was included for next stage.

---

[14] "The extent to which the people, interventions, and outcome measures are similar to those of interest." [8]

Other limitation to consider is the fact of this review has an "*inaccessible list*" of 26 papers that *apparently*, have some relevance for this research, but could not be assessed to confirm this supposition. Meanwhile, this number of papers can be considered low (2%) if we consider the amount of 1272 input papers of the stage 3. In other words, this aspect can be considered part of the consistency of the strength of evidence found by this review, that despite being an aspect that cannot be ignored, it does not mischaracterize the representativeness the final findings.

All along the pilot of the **data extraction** procedure (stage 4), we found that various papers lacked enough details about the study design and related findings. Therefore, in the beginning, we diverged too much in what indeed we extracted. Directed toward reduce the bias, the data from the entire set of studies at this stage were extracted by two researchers separately (on each pair), conforming to the predetermined extraction form [15], and this information were gathered in a database through the tool developed during this review. After that, the researchers from each pair carried out a set of meetings to select the *final extracted data* by consensus, using another interface of the same tool. Moreover, we realized that the extraction process was constantly hampered by the manner some studies were disclosed. Lamentably, sometimes the documentation procedure cannot be satisfactorily carried out using the extraction form, due several studies lacked sufficient information to do that. Hence, there is a chance that the extraction procedure might have conducted to some imprecision in the data. Anyhow, we believe that because the methodological rigor followed by this review, the universe set of studies obtained, is at least, *representative sampling from the phenomena under study*.

## 5. CONCLUSIONS

This paper brings the following **main contributions**: (1) advance in the state of art of agile governance, providing a mapping of findings organized in four major groups and 16 categories, which can be figured by relevance and convergence; (2) the characterization of agile governance as a nascent multidisciplinary socio-technical phenomena positioned in a chaordic range between the innovation and emergent practices from agile philosophy and the status quo of the best practices employed and demanded by the governance issues, focused on organizational performance and competitiveness; and, (3) a new and convergent definition for agile governance, six meta-principles, the concept of a dynamic repository for the knowledge related to governance topic (GBOK) and some directions for research and practice.

As **additional contributions**, we improved and complemented the methodological approach on which was based this review (see section 3): (i) upgrading and adding new procedures to carry out a systematic review; (ii) as well as developing a tool to support this research in a geographically distributed manner; and, (iii) adopting quantitative procedures to develop a qualitative analysis of the evidences. We believe that this approach can, at least, inspire researchers and practitioners in future qualitative researches and systematic reviews.

An evident finding of the review is that agile governance is a nascent, wide and multidisciplinary domain, focused on organizational performance and competitiveness that needs to be more intensively studied and might have its boundaries better defined in **future works**. Absolutely, we can realize that there is a research backlog with topics that need to be tackled. Naturally, researchers and practitioners should work together to define a reciprocal research agenda for new research and practice on this domain, as well as to get closer to the wide spectrum of interest and application identified by this review. Furthermore, due the amplitude of the agile governance domain and the large number of findings identified by this systematic review, we consider the publication of other relevant aspects of this review as a **future work**.

Finally, the authors believe that not only software development industry, manufacturing industry, and IT industry, but also whole business industry, will **benefit** with these results. Due once the agile governance phenomena are better understood in their essence, starting by its concept and application, as well as how it evolved over the time; become possible, in a second stage, map their constructs, mediators, moderators and disturbing factors from those phenomena in order to help organizations to achieve better results in their application: reducing cost and time, increasing the quality and success rate of their practice. As a consequence, improving the competitiveness of governments and companies through the improvement of their governance and management shall result in significant economic returns.

International Journal of Computer Science & Information Technology (IJCSIT) Vol 6, No 5, October 2014

## ACKNOWLEDGEMENTS

We applied the SDC approach for the sequence of authors [36]. The authors acknowledge to CAPES, Brazil's Science without Borders Program, CNPq, ATI-PE and SERPRO by the research support. Thanks to all the researchers who participated in this systematic review in any of the phases, of which we cite: *Derick Hsieh, Guilherme Dias da Silva, Renata Maria Andrade do Nascimento, Reynaldo Fabrinny Rodrigues Tibúrcio, Jirlaine Fonseca de Oliveira, Edmundo Rodrigues da Silva Porto Neto and Fabio Santos Batista*. We are grateful to *Tatiana Bittencourt Gouveia* and Prof. Dr. *Fabio Q. B. da Silva* for having kindly provided their template of the research protocol, as an inspiration for elaboration of the protocol of this review. Special thanks to *Luciano José de Farias Silva* and *Daniel de Andrade Penaforte* for their valuable contributions. The research team thank to the authors and publishers that have made available their papers for assessment, when asked about it.

## APPENDIX A. STUDIES INCLUDED IN THE REVIEW

As a matter of length, we are depicting in this section only studies directly cited in the text of this paper. See the entire list, and references, about the studies included in this review at: http://www.agilegovernance.org/slr-ag-up-to-2013/studies.

## Authors

As a matter of length, the authors' bio is available at: http://www.agilegovernance.org/slr-ag-up-to-2013/authors.